# Two-step model versus one-step model of the inter-polarization conversion and statistics of CdSe/ZnSe quantum dot elongations

A.V. Koudinov, B.R. Namozov and Yu.G. Kusrayev
A.F. Ioffe Physico-Technical Institute of RAS, 194021 St.-Petersburg, Russia

S. Lee
Department of Physics, Korea University, Seoul 136-701, Korea

M. Dobrowolska and J. K. Furdyna
Department of Physics, University of Notre Dame, Notre Dame, IN 46556

The magneto-optical inter-polarization conversions by a layer of quantum dots have been investigated. Various types of polarization response of the sample were observed as a function of external magnetic field and of the orientation of the sample. The full set of experimental dependences is analyzed in terms of a one-step and a two-step model of spin evolution. The angular distribution of the quantum dots over the directions of elongation in the plane of the sample is taken into account in terms of the two models, and the model predictions are compared with experimental observations.

**Introduction**

Self-assembled CdSe/ZnSe quantum dot (QD) samples have long been the subject of intense research, with many articles devoted to their radiative and spin-related properties. Most recently considerable attention has been given to various schemes of spin manipulation in systems of this type in light of its relevance to quantum data processing algorithms.[1]

Fine structure of the exciton ground state is typical for electrically neutral self-assembled CdSe/ZnSe QDs and related low-dimensional systems. The quartet of the exciton states is split by isotropic exchange interaction into a doublet of optically inactive ("dark") excitons, characterized by the projections of the angular momentum $\pm 2$, and a doublet of optically active ("bright") exciton states with angular momentum projections of $\pm 1$.[2] In the case of CdSe/ZnSe QDs a typical energy separation between the doublets is about 2 meV. Since the QD states are anisotropic in the plane of the QD layer, the bright excitons doublet is further split by an anisotropic exchange interaction. This results in two distinct energy levels, coupled to light with orthogonal linear polarization.

We now consider the behavior of this system in an external magnetic field. If the field is applied along the growth axis $z$, the selection rule characteristics of the bright excitons doublet changes, in that the pure linear polarizations of the dipole transitions associated with the QD anisotropy gradually acquire an admixture of circular polarization. A direct manifestation of the change of symmetry of optical transitions resulting from this process is the inter-polarization conversion that occurs, e.g., when the optical excitation by a linearly polarized light leads to an admixture of circular polarization of the emission. Such polarization conversion occurs in the range of applied fields where the Zeeman splitting of the bright excitons is comparable to the anisotropic exchange splitting (whose magnitude is estimated to be *ca.* 0.2 meV[3]).

Polarization conversions of the type described above were first observed in bulk GaSe[4], and were subsequently found in semiconductor nanostructures such as GaAs/GaAlAs [5,6,7]. An important advance in this area was made by the application of the pseudospin method to the problem of polarization conversion.[8] This method was then applied to the analysis of experimental results observed on InAlAs/AlGaAs,[9] InAs/GaAs,[10] and CdSe/ZnSe[11] QDs in an applied magnetic field. Recently, conversions of linear-to-circular and circular-to-linear polarization were also observed for CdSe/ZnSe QDs in the absence of magnetic field.[12]

In Ref. [11], the experimental results were interpreted in terms of a two-step model. In this model one assumes a rather long lifetime of the exciton in its ground state, in agreement with time-resolved spectroscopy data. In that work it was concluded that there is a directional scatter of the polarization axes of individual CdSe/ZnSe QDs. To account for this scatter, the analysis in Ref. [11] introduced a consciously crude assumption that there were *two* families of QDs in the QD ensemble, one with QDs elongated along the

{110}-type directions, and the other along the {100}-type directions. In Ref. [12] and other articles the analysis was carried out in terms of a simple one-step model of polarization conversion .

In this paper we present a complete set of data on optical polarization response of excitons in anisotropic QDs, including results for all types of polarization conversions and their dependences on the orientation of the sample. The anisotropy of linear-to-linear polarization conversion at zero magnetic field has been observed for the first time. The linear-to-circular and circular-to-linear conversion is induced by an applied magnetic field, and vanishes at zero field. Using the pseudospin approach, we have calculated the magnetic field- and the angular dependences of all possible types of polarization response in terms of the two-step and the one-step model, and the results were critically compared to the experimental data. The distribution of the QDs over the directions of their elongation in the plane of the sample is taken into account in both the models. The parameters of the distribution are then determined by comparison of the measured results with the calculated two-step and one-step pictures of the conversion phenomenon.

**Calculation**

The pseudospin method was first applied to the problem of polarization conversion in Ref. [8]. Its applicability to this problem is restricted by the condition that the bright excitons doublet should be isolated from other states (essentially by the condition $\delta_1 \ll \delta_0$, where $\delta_1$ and $\delta_0$ are the characteristic energies of the anisotropic and isotropic exchange interactions, respectively). It is fortunate that in practice this condition is fulfilled in typical nanostructures. At the same time, definite advantages of the method reside in its clarity and the transparency of the obtained results, as compared to rigorous treatments using the spin density matrix technique. The possibility to obtain results rather easily *and in analytic form* is invaluable in the analysis of the rather complicated physical scenarios encountered in real experimental situations.

In the experimental part of the present work, the measured quantities are given in the form of three degrees of polarization, defined by

$$\rho_C = \frac{I_+ - I_-}{I_+ + I_-}, \quad \rho_L = \frac{I_\xi - I_\eta}{I_\xi + I_\eta}, \quad \rho_{L'} = \frac{I_{\xi'} - I_{\eta'}}{I_{\xi'} + I_{\eta'}}, \tag{1}$$

where $I_\alpha$ represents the light intensity in the $\alpha$ polarization, the subscript $\alpha$ denoting the right- and left-hand circular polarizations (+ and –); two orthogonal linear polarizations $\xi$ and $\eta$ (which for convenience we will define as "vertical" and "horizontal") ; and, finally, the linear polarizations $\xi'$ and $\eta'$ rotated by 45° with respect to $\xi$ and $\eta$. The parameters defined by Eqs. (1) completely characterize any polarization of light, serving as representations of the three Stokes polarization parameters. The objective is to calculate the degree to which the three components of polarization are present in the emitted photoluminescence when the system is excited by either circularly ($C$) or linearly polarized light ($L$). In the case of excitation by linearly polarized light a reasonable choice for the direction of polarization $L$ is along $\xi$. Thus we shall look for six different types of polarization response: $CC$ (optical orientation), $CL$ and $CL'$ (circular-to-linear conversion), $LL$ (optical alignment), $LL'$ (linear-to-linear conversion) and $LC$ (linear-to-circular conversion).

The essence of the pseudospin method is based on the fact that the polarization properties of the bright exciton emission are associated with the direction of the vector of exciton pseudospin $1/2$ in an effective space. The anisotropic exchange splitting of the bright exciton doublet that arises from the anisotropy of exciton potential in the plane of the sample can then be represented as an effective magnetic field (the "exchange field") acting on the pseudospin. The pseudospin space $OXYZ$ is defined such that the $Z$ axis coincides with the growth axis $z$ of the structure, and the $XY$ plane coincides with the $xy$ (i.e., the layer) plane. Note that an angle between two arbitrary directions in the $xy$ plane corresponds to twice the angle in the $XY$ plane. The three degrees of polarization defined by Eqs.(1) correspond to twice the projections of the pseudospin on the respective directions: $\rho_C = 2S_Z$, $\rho_L = 2S_X$, $\rho_{L'} = 2S_Y$, indicating the analogy of the pseudospin space and the Poincare sphere.[13] The external magnetic field is directed along the $Z$ axis, while the exchange field is always in the $XY$ plane. It is important to note that in the real $Oxyz$ space the in-plane anisotropy of the potential resulting from, e.g., a compression along $x$ is physically the same as a compression along the opposite ($\bar{x}$) direction. This allows the equatorial $XY$ plane in the pseudospin $OXYZ$ space to contain all physically nonequivalent orientations of the exchange field.

Consider the precession of the pseudospin $\vec{S}$ in a total "magnetic field" (which includes the applied and exchange fields). We shall represent this field by the vector $\vec{\Omega}$, whose value equals the Larmor frequency produced by the total field, and whose direction coincides with the direction of that field. It can be shown that the corresponding Bloch equation,

$$\dot{\vec{S}} = \vec{\Omega} \times \vec{S}, \tag{2}$$

with the initial condition

$$\vec{S}(t=0) = \vec{S}_0, \tag{3}$$

has the solution

$$\vec{S}(t) = \vec{e}_\Omega (\vec{e}_\Omega \vec{S}_0) + [\vec{e}_\Omega \times \vec{S}_0] \sin \Omega t + (\vec{S}_0 - \vec{e}_\Omega (\vec{e}_\Omega \vec{S}_0)) \cos \Omega t, \tag{4}$$

where $\vec{e}_\Omega = \vec{\Omega}/\Omega$ is the unit vector along the direction $\vec{\Omega}$.

In what follows we shall need the pseudospin averaged over the distribution of the lifetimes in a given state $W(t) = \tau^{-1} \exp(-t/\tau)$:

$$\langle \vec{S} \rangle_\tau = \vec{e}_\Omega (\vec{e}_\Omega \vec{S}_0) + [\vec{e}_\Omega \times \vec{S}_0] \frac{\Omega \tau}{1 + \Omega^2 \tau^2} + (\vec{S}_0 - \vec{e}_\Omega (\vec{e}_\Omega \vec{S}_0)) \frac{1}{1 + \Omega^2 \tau^2}. \qquad (5)$$

In Ref. [11] the following model was used for interpretation of the experiments on the optical orientation and optical alignment of excitons in the ensamble of CdSe/ZnSe QDs. The laser excitation at an energy somewhat above the excitonic resonance was assumed to create excitons in a short-lifetime excited state, from which they rapidly escape to the lower-lying ground state, where they spend the major part of their lifetime before recombining via photon emission (Fig.1(*a*)). The values of the exchange fields in the excited (upper) and the ground (lower) states may be different, and we shall denote the corresponding Larmor frequencies by $\Omega_{ex}$ and $\omega_{ex}$. However, the directions of these fields were assumed to coincide, which is justified because both these states belong to the same (anisotropic) QD. Estimates based on typical lifetimes and on the values of the anisotropic exchange splitting $\delta_1$ have shown that in the lower state $\omega_{ex} \tau \gg 1$, i.e., even in zero external field the gain in the phase of the precession $\Omega \tau$ is large, so that only the projection of the initial pseudospin on the direction of the total field (the first term in Eq.(5)) is conserved until the moment of recombination. Thus the evolution of the pseudospin in the lower state is reduced to its projection along the direction of the field, while in the short-lifetime upper state all three terms in Eq.(5) may be essential.

In the present paper we shall consistently calculate polarization responses in terms of the described two-step model. For comparison, we shall in parallel calculate the same responses within the simple one-step model, in which all the evolution of the pseudospin occurs only in the ground exciton state (Fig.1(*b*)).

For our purposes it will be convenient to slightly modify the approach of the previous papers by defining the direction $X$ with respect to the "vertical" direction $\xi$ in the laboratory reference frame, rather than to one of the high-symmetry axes of the crystal.[8,9] With this choice excitation by the circularly polarized light ($C$) corresponds to

$$\vec{S}_0 = (0, 0, 1/2), \qquad (6)$$

and by linearly polarized light ($L$), to

$$\vec{S}_0 = (1/2, 0, 0). \qquad (7)$$

We assume that all QDs have the same elongated (e.g., elliptical) shape in the sample plane, which determines the anisotropic exchange interaction. For the total field in the upper and lower states, one can write, respectively,

$$\vec{\Omega} = (\Omega_{ex} \cos \Phi, \Omega_{ex} \sin \Phi, \Omega_B), \qquad (8)$$

$$\vec{\omega} = (\omega_{ex}\cos\Phi, \omega_{ex}\sin\Phi, \Omega_B), \qquad (9)$$

where $\Omega_B$ denotes the pseudospin precession frequency in the external magnetic field, and $\Phi$ is the angle (in the pseudospin space) between the direction $X$ and the direction of the exchange field in a given QD. Then, by considering Eq.(5) as an expression for the mean spin in the upper state and projecting it onto the direction of the total field in the lower state,

$$\vec{S}_{fin} = \vec{e}_\omega \left( \vec{e}_\omega \langle \vec{S} \rangle_\tau \right), \qquad (10)$$

one obtains via Eqs. (8) and (9) the following results. For circularly polarized excitation, i.e., with the initial condition Eq.(6),

$$\begin{pmatrix} CL \\ CL' \\ CC \end{pmatrix} = \begin{pmatrix} \omega_{ex}\cos 2\phi \\ \omega_{ex}\sin 2\phi \\ \Omega_B \end{pmatrix} \frac{\Omega_B}{\omega_{ex}^2 + \Omega_B^2} \cdot \frac{1 + (\omega_{ex}\Omega_{ex} + \Omega_B^2)\tau^2}{1 + (\Omega_{ex}^2 + \Omega_B^2)\tau^2}, \qquad (11)$$

while for the linearly polarized excitation, with Eq.(7),

$$\begin{pmatrix} LL \\ LL' \\ LC \end{pmatrix} = \begin{pmatrix} \omega_{ex}\cos 2\phi \\ \omega_{ex}\sin 2\phi \\ \Omega_B \end{pmatrix} \frac{\left(\omega_{ex} + \Omega_{ex}(\omega_{ex}\Omega_{ex} + \Omega_B^2)\tau^2\right)\cos 2\phi + \Omega_B(\omega_{ex} - \Omega_{ex})\tau\sin 2\phi}{(\omega_{ex}^2 + \Omega_B^2)(1 + (\Omega_{ex}^2 + \Omega_B^2)\tau^2)}. \qquad (12)$$

In both these expressions we performed substitution $\Phi \to 2\phi$, which means the transfer from the pseudospin space to the real space. Equations (11) and (12) describe all six forms of the polarization response of one QD as a function of the value of the applied magnetic field $\Omega_B$ and the angle $\phi$ between the "vertical" direction and the direction of elongation (the long axis) of a given QD. These equations may also be considered as a final result for the ensemble in which all QDs (or indeed all states in a quantum well or a superlattice) are elongated in a single direction. Analogous formulas obtained by the one-step model are given in the Appendix.

We now analyze the results obtained via this approach. One can see from Eq.(11) that for circularly polarized excitation the conversions $CL$ and $CL'$ are strictly odd in the magnetic field for any $\phi$, while the optical orientation $CC$ is strictly even in the field. In contrast, as can be seen from Eq.(12), for linearly polarized excitation none of the responses show a strict parity in the field. Furthermore, one can easily see that if the angle $\phi$ takes all values with equal probability, then in Eq.(11) only the $CC$ response (which is independent of $\phi$) will remain nonzero, while in Eq.(12) the non-vanishing responses will be $LL$ (the part containing $\cos^2 2\phi$) and $LL'$ (the part containing $\sin^2 2\phi$). Note that with this assumption the conversion $LL'$ differs from zero as long as the values of the exchange field in the upper and lower states ($\Omega_{ex}$ and $\omega_{ex}$) are different. Other nonzero responses in the model under consideration are possible only if one assumes an anisotropy in the angular distribution of the QD elongations.

Relying on the qualiatative conclusions of many papers and on our recent experience with the analysis of the angular harmonics of magnetic-field-induces linear polarization,[14] we choose to model the distribution of the orientations of QD long axes in the sample plane by the function

$$R(\psi) = \frac{1}{2\pi}\left(1 + \frac{a}{1+a+b}\cos 2\psi + \frac{b}{1+a+b}\cos 4\psi\right), \qquad (13)$$

where $\psi$ is the angle (in the real space) measured from the [110] direction, and parameters $a$ and $b$ are positive. Here the first term characterizes the isotropic part of the distribution; the second term indicates the preference of the major axis for the [110] orientation over [1$\bar{1}$0] (a situation that is commonly encountered in typical nanostructures, for reasons that are not yet fully understood); and the third indicates the preference of QD orientation for {110}-type directions over {100} (which is a reflection of the cubic symmetry of the crystal forming the nanostructure).

Averaging Eqs.(11) and (12) over the distribution given by Eq. (13) results in

$$\begin{pmatrix} CL \\ CL' \\ CC \end{pmatrix} = \begin{pmatrix} \omega_{ex}\dfrac{a}{1+a+b}\dfrac{\cos 2\varphi}{2} \\ \omega_{ex}\dfrac{a}{1+a+b}\dfrac{\sin 2\varphi}{2} \\ \Omega_B \end{pmatrix} \dfrac{\Omega_B}{\omega_{ex}^2 + \Omega_B^2} \cdot \dfrac{1+(\omega_{ex}\Omega_{ex}+\Omega_B^2)\tau^2}{1+(\Omega_{ex}^2+\Omega_B^2)\tau^2}, \qquad (14)$$

$$\begin{pmatrix} LL \\ LL' \\ LC \end{pmatrix} = \begin{pmatrix} \omega_{ex}\left(\dfrac{1}{2}+\dfrac{b}{1+a+b}\dfrac{\cos 4\varphi}{4}\right) \\ \omega_{ex}\dfrac{b}{1+a+b}\dfrac{\sin 4\varphi}{4} \\ \Omega_B\dfrac{a}{1+a+b}\dfrac{\cos 2\varphi}{2} \end{pmatrix} \dfrac{\omega_{ex}+\Omega_{ex}(\omega_{ex}\Omega_{ex}+\Omega_B^2)\tau^2}{(\omega_{ex}^2+\Omega_B^2)(1+(\Omega_{ex}^2+\Omega_B^2)\tau^2)} +$$

$$+ \begin{pmatrix} \omega_{ex}\dfrac{b}{1+a+b}\dfrac{\sin 4\varphi}{4} \\ \omega_{ex}\left(\dfrac{1}{2}-\dfrac{b}{1+a+b}\dfrac{\cos 4\varphi}{4}\right) \\ \Omega_B\dfrac{a}{1+a+b}\dfrac{\sin 2\varphi}{2} \end{pmatrix} \dfrac{\Omega_B(\omega_{ex}-\Omega_{ex})\tau}{(\omega_{ex}^2+\Omega_B^2)(1+(\Omega_{ex}^2+\Omega_B^2)\tau^2)} \qquad (15)$$

Equations (14) and (15) describe the entire set of polarization responses of the ensemble of identical anisotropic QDs, whose in-plane orientations are distributed according to Eq.

(13), as a function of applied magnetic field $\Omega_B$ and of the angle $\varphi$ between the "vertical" direction and the [110] axis of the sample. Analogous expressions obtained by the one-step model are given in the Appendix.

In Eqs. (11), (12), (14) and (15) the relaxation of the pseudospin was not taken into account. This is justified by the fact that in QDs typical relaxation times are longer than the excitonic radiative lifetimes. In the upper state one is even more justified to neglect pseudospin relaxation, since in that case the pseudospin lifetime $\tau$ is even shorter (by several orders of magnitude) due to rapid escape to the lower state. Taking the relaxation in the lower state into account would lead to the appearance on the right-hand side of Eqs. (11), (12), (14) and (15) of a common prefactor $T_s / \tau_{rad}$, where $T_s^{-1} = \tau_s^{-1} + \tau_{rad}^{-1}$, and $\tau_s$ and $\tau_{rad}$ are, respectively, the pseudospin relaxation time and the lifetime of the lower state.

**Experimental results and discussion**

Figure 2 shows the photoluminescence (PL) spectrum of the QD ensemble[15] excited at the energy 2.330 eV ($\lambda = 532$ nm), i.e., slightly above the main excitonic transition (1$e$-1$hh$ in common notation), together with the spectrum of optical alignment (i.e., the degree of linear polarization of the PL observed for linearly polarized excitation). Superimposed on the inhomogeneously broadened PL line one can see distinct features of the optical phonon cascade. The phonon cascade is more pronounced in the polarization spectrum.

As is well known, the phonon structure is due to more efficient excitation of those QDs for which the energy of the main excitonic transition differs from the energy of the exciting photons by an integer number of optical phonon energies. It then appears natural that those states which are populated faster also better preserve the "spin memory", thus reproducing the polarization of the incident laser beam. The same idea allows us to understand the overall decay of the polarization of the signal seen in Fig. 2 as the energy decreases – spectral diffusion requires time. However, an additional factor can also contribute to the decrease of the polarization at the low-energy end: the increasing contribution from the emission of trions, i.e., singly charged excitons. Independent of the trion type – whether positively or negatively charged – the singlet state of the two like particles in the trion (two electrons or two holes) will destroy the correlation between electron and hole spins, thus preventing optical alignment.

For the study of the inter-polarization conversion we have chosen 2.246 eV for data acquisition (shown by the arrow in Fig.2). This choice was convenient, because the significant polarization at this energy is also accompanied by high PL intensity, which allowed efficient accumulation of statistics in the polarization measurements.

Experimental results for the inter-polarization conversions are presented in Figs. 3 and 4. Of the twelve graphs in the figures, only one (Fig.4(*d*)) has no relation to spin memory.

That latter case shows the angular dependence of the linear polarization of the PL taken with non-polarized excitation at zero external field. The polarization of that kind (the "built-in polarization") is an intrinsic property of the sample, revealing unambiguously its in-plane anisotropy. The built-in polarization is frequently observed in quantum well samples, and was also found in CdSe/ZnSe QDs in Ref. [12]. As a rule (this is also the case in the present study), it reveals non-equivalence of the directions [110] and [1$\bar{1}$0], although for individual QDs the predominant directions can differ for different QDs.[16,17,18,19] As for the microscopic origin of the built-in polarization, in case of neutral excitons, it can be due to thermalization – the preferential population of the lower-energy levels of the bright excitonic doublets, which is a consequence of the Boltzmann distribution. While experiments on single CdSe/ZnSe QDs have shown that the thermalization is weak (because the exciton lifetime is too short),[16,17] the observed signal of the built-in polarization is also not large. In our case, the idea of thermalization is supported by the onset of "thermal" circular polarization in the external magnetic field at excitation far above the exciton line or at quasi-resonant excitation by the non-polarized light (not shown). Other possible mechanism of the built-in polarization is the mixing of states of the heavy-hole and light-hole subbands caused by in-plane anisopropy of the QD. This mechanism determines the built-in polarization in trions, but can also apply to the neutral excitons. In principle both mechanisms can contribute to the amplitude of the built-in polarization.[20]

As to the direction of the built-in polarization, this is obviously fixed with respect to the crystal axes, the polarization plane being parallel to the [110] direction. The symmetry of the observed angular dependence – a pure second angular harmonic – is a simple result of rotation of the sample with respect to the laboratory reference frame (with respect to which we define the degree of polarization). The same symmetry of the signal would have been produced by a lamp and a polarizer rotated in front of the analysing optics.

The remaining 11 graphs in Figs. 3 and 4 are related to the spin memory effects and should be compared with the models treated in the theory section. The plotted experimental data have been corrected for accompanying extraneous effects. In particular, in the magnetic field dependences in Fig. 3 the experimental values of the "thermal" circular polarization obtained with the non-polarized excitation ($NC$, not shown) have been subtracted from the experimental values of polarization as measured in the $CC$ and $LC$ configurations. The corrected results are shown in Figs. 3($a$) and 3($h$). The thermal polarization $NC$ is not large (about 2% at the largest field value), its field dependence being odd in the field. In an analogous way, the signal of circular dichroism of absorption (which turned out to be small in value) has been subtracted from the data as measured in the $CL$ configuration. From the angular dependences obtained in configurations $LL$ and $LL'$, the built-in polarization shown in Fig.4($d$) has been subtracted, and the result is shown in Fig.4($a$) and Fig.4($b$).

Of the magnetic field dependences depicted in Fig.3, the largest signals are observed in the optical orientation (Fig.3($a$)) and the optical alignment (shown for two orientations of the crystal, Figs. 3($b$) and 3($c$)). The values of the inter-polarization conversions (Figs.3($d$) to 3($h$)) lie in the range of $\pm 5\%$. The curves of all the conversions have a standard S-like shape,[7,8,9] and they all look like odd functions of the field – all but the conversion $LL'$ for the angle $\varphi = 22.5°$ (Fig.3(*f*)). In the latter case a nonzero value of the polarization response is observed at zero field – the S-shaped curve is shifted upwards and somewhat distorted. The angular dependences in Fig.4 can help to make this result clear.

Figure 4(*a*) shows the angular dependence of the optical alignment effect $LL$ at zero extenal field. In the polar frame, it looks like a slightly dimpled circle, that is, the optical alignment is nearly isotropic. This means that a model with aligned long axes of the QDs does not correspond to our experimental situation, nor does the model with all the QDs elongated along some high-symmetry directions (e.g., along {110}-type axes), because in both these cases one would expect the $LL$ signal to vanish at some positions of the crystal, e.g., when the [100] axis is parallel to the "vertical" direction (i.e., at $\varphi = 45°$). Nevertheless, an anisotropic component is present in the angular scan of Fig. 4(*a*), which can be expressed by a fourth angular harmonic. The alignment is maximum along the {110}-type directions ($\varphi = 0°, 90°, 180°, 270°$) and minimum along the {100}-type directions ($\varphi = 45°$, etc.), the fact related to the angular scatter of the long axes of the QDs.

The anisotropy is much more pronounced in the angular scan of the $LL'$ conversion, as shown in Fig. 4(*b*). Here one clearly sees the pure fourth harmonic, without any isotropic component. The angular scan of $LL'$ is a sign-changing function, with a maximum at $\varphi = 22.5°$, etc. Note that the opposite sign of the $LL'$ response at opposite inclinations of the crystal (e.g., from $\varphi = 0°$ toward positive and negative angles) does not indicate an asymmetry of the physical picture for the two directions of rotation, but simply results from the choice of sign in the definition of the polarization $L'$.

The dependences like those depicted in Figs. 4(*a*) and 4(*b*) have not been observed before. What do they mean from the physical point of view? For linearly polarized excitation and $\varphi$-rotation of the sample, the value of the total degree of linear polarization of the PL, $\rho_{tot} = \left(\rho_L^2 + \rho_{L'}^2\right)^{1/2}$, oscillates between about $20-25\%$ with a periodicity of 90-degrees (maximum polarization at $\varphi = 0°$, etc.). The direction of the total polarization also oscillates with a 90-degree periodicity, being inclined toward and away from the "vertical" direction of polarization of the incident light – as if the PL polarization plane were attracted to the nearest {110}-type direction. For the sample under study, the magnitude of these swings amounts to about $3°$.

In Fig.4*c*, the angular scan of the polarization conversion $LC$ is presented. To the best of our experimental accuracy, and in contrast with the results obtained in Ref. [12], no $LC$ conversion is observed at zero field. A similar result is obtained for the $CL$ conversion; however, here one should take care to subtract the admixed built-in polarization from the measured data. The reasons for the disagreement between our results and the observations of Ref. [12] are discussed below.

Let us now turn to testing of the two-step and one-step models of polarization conversion. The two-step model, expressed by Eqs. (14) and (15), is controlled by the following parameters: $\omega_{ex}$, $\Omega_{ex}$, $\tau$, $a$, and $b$. For the sixth parameter, we have chosen the common prefactor $D$ entering the right-hand side of Eqs. (14), (15). As mentioned in the theory section, such a common prefactor is naturally obtained by assuming the relaxation of the pseudospin in the lower level. It then takes the form $T_s / \tau_{rad}$, but we prefer to express it as $D$ in order to emphasize that the prefactor phenomenologically accounts for *all* dissipative mechanisms that reduce the pseudospin in the final state – e.g., possible losses due to the imperfections of selection rules.

The results of simulation of the eleven experimental dependences by this six-parameter model are shown as solid curves in Figs. 3 and 4 (*a*) – (*c*). These curves have been calculated with the following values of the parameters:[21] $\omega_{ex} = 0.3\,\text{meV}$, $\Omega_{ex} = 2.1\,\text{meV}$, $\tau = 0.2\,\text{ps}$, $a = 0.8$, $b = 0.5$ and $D = 0.46$. Comparison with experimental results allows us to evaluate four parameters with a high degree of reliability: $\omega_{ex}$, $a$, $b$ and $D$. The choice of the values for $\Omega_{ex}$ and $\tau$ is, on the other hand, somewhat arbitrary, since an increase of either of these parameters requires a corresponding decrease of the other. This arises from the fact that the *product* $\Omega_{ex}\tau$ represents the total phase gain in the upper state during its lifetime. While we have not been able to find good agreement with experiment at the condition $\Omega_{ex} \sim \omega_{ex}$, the simulation works very well when the value $\Omega_{ex}$ is substantially larger than $\omega_{ex}$ (say, by 5 times or more). Moreover, good agreement is obtained if one chooses $\Omega_{ex}$ and $\omega_{ex}$ to be of opposite signs (the opposite signs imply opposite sequence of the pairs of levels in the upper state and lower state doublets) when $\Omega_{ex}$ is about two or more times larger than $\omega_{ex}$. Here one should note that the excited states of the exciton in a QD can have either sequence of levels.[22] Whether $\Omega_{ex}$ is positive or negative, in the case of QDs under study its absolute value appears to be several times larger than that of $\omega_{ex}$, while $\tau$ lies in the sub-picosecond range. A larger strength of the anisotropic exchange interaction in the excited exciton state was reported for the CdSe/ZnSe QDs in Ref. [23].

The frequently-used one-step model of polarization conversion[8,9,10,12] is controlled, in our approach (Eqs. (A3), (A4)), by five parameters: $\Omega_{ex1}$, $\tau$, $a$, $b$ and $D$. The result of simulation of the experimental dependence is shown in Figs. 3 and 4(*a*) – (*c*) by dashed curves, calculated with the following values: $\Omega_{ex1} = 0.32\,\text{meV}$, $\tau = 10\,\text{ps}$, $a = 0.8$, $b = 0.53$ and $D = 0.44$. We note that the value $\Omega_{ex1}$ turns out to be very close to the value $\omega_{ex}$ determined in the two-step model, and the values $a$, $b$ and $D$ for the two models almost coincide. In the one-step model, there is practically no freedom in the choice of parameters, so that all the five values are determined quite rigidly. The first impression is that the overall picture of the phenomenon is described equally well by the one-step model. However, some features of the one-step model are far from compatible with the physical situation.

Firstly, the value $\tau \sim 10$ ps given by this model looks unrealistic, since time-resolved spectroscopy experiments on CdSe/ZnSe QDs give exciton lifetimes of about $\tau_{rad} = 300 - 500$ ps.[24,25,26] One could formally try to identify the value of 10 ps obtained in the one-step fitting process with the pseudospin relaxation time. However, as mentioned in the Appendix, if the relaxation were taken into account, the prefactor $T_s / \tau_{rad} \sim 10 ps / \tau_{rad} \leq 0.03$ would have appeared in the right-hand sides of Eqs. (A3) and (A4), in obvious contradiction with the obtained value of $D = 0.42$. We must thus conclude that the assumption of the rapid relaxation of the pseudospin (in the 10 ps range) is incompatible with the experimentally observed rather large values of polarization $\sim 50\%$.

Secondly, some difficulties arise in the description of the magnetic field dependences of the $LC$ and $CL$ conversion (Fig.3(*d*) and 3(*h*)). The point is that in the one-step model the two phenomena are described by exactly the same expression, while the experiment yields the S-shaped curves with somewhat different magnitudes. Different amplitudes of $LC$ and $CL$ conversions were found in Refs. [8,10], and in Ref. [8] this difference was interpreted in terms of the "cascade" process, i.e., involving higher-lying excitonic states. Indeed, in can be shown using Eqs. (14), (15) that the ratio of the two conversions within the two-step model is determined by the difference in the values of $\Omega_{ex}$ and $\omega_{ex}$. Specifically, at $\Omega_{ex} > \omega_{ex}$, the $LC$ conversion has larger amplitude than the $CL$ conversion, which agrees with experimental data.

Thirdly, the one-step model predicts the $LC$ and $CL$ conversions in the absence of the magnetic field. In the configuration $LC$, the maximum value of the circularly polarized response should be achieved with the polarization of the incident linearly polarized light parallel to the {100}-type directions. This is just the behavior observed in Ref. [12]. In the $CL$ configuration the linear polarization should appear directed along [100] or along [010], depending on the handedness of the incident $C$ excitation.[27] The same symmetry of the signal would have been obtained if there were, in place of the sample, a quarter-wave plate with two principal directions along [110] and [1$\bar{1}$0].

These effects should have manifested themselves in the angular scans of $LC$ and $CL$ in the form of patterns having the symmetry of the second angular harmonic. However, our experiments did not reveal such $2\varphi$-contributions (see Fig.4(*c*)). While with the set of parameters determined for the one-step model the expected amplitudes of the $2\varphi$-contributions is indeed small $\sim 1.5\%$ (thin solid line in Fig.4(*c*)), one can see from Fig. 4(*d*) that the built-in polarization, which has a similar symmetry with an amplitude twice as small, was recorded quite clearly in our experiments. Here the absence of the conversions $LC$ and $CL$ at zero field ought to be considered as a serious argument against the one-step model. This therefore provides strong support in favor of the two-step model, where zero-field conversions $LC$ and $CL$ are strictly absent.

We cannot ignore the fact that in Ref. [12] the conversions *LC* and *CL* at zero field in CdSe/ZnSe QDs have been experimentally observed. It is possible, however, that the difference in the results of the present paper and of Ref. [12] is related to different experimental conditions or to a difference in the QD morphology. In Ref. [12] the excitation was closer to the excitonic resonance, and the polarization data were taken at the sharp 1LO-phonon replica of the laser line. The zero-field conversion was observed at the 1LO-line, but at other detection energies it was not found,[28] similar to our result. One can thus suggest that the presence of the zero-field conversion phenomenon in Ref. [12] is related to the peculiarities of the radiative process responsible for the 1LO-line in the spectrum.

In Fig.5 we plot the reconstructed distribution of the QDs under study over the directions of the elongation axes. The deduced values of parameters *a* and *b* controlling the shape of the distribution are very close for the two polarization conversion models, so the result is almost model-independent. Admittedly the choice of the angular distribution in the form of Eq. (13) is somewhat arbitrary, and one might have chosen another function obeying the same overall symmetry requirements. It is remarkable, however, that then the agreement between the calculation and the experiment would become worse. This can be conveniently illustrated by the example of the angular scans of the *LL* and *LL'* responses. For instance, one can see from Eq. (15) that at zero magnetic field ($\Omega_B = 0$) the *LL'* response contains only the fourth angular harmonic ($\sin 4\varphi$), while for a single QD it contains (see Eq.(12)) the second ($\sin 2\phi$) and the fourth ($\sin 2\phi \cos 2\phi$) harmonics. The absence of the second harmonic in the averaged responses *LL* and *LL'* according to Eq. (15) is not caused by any fundamental symmetry reasons, being just a property of the particular form of the angular distribution function, Eq. (13) – a consequence of the wide "petals" in the pattern shown in Fig. 5. With a different choice of angular function the second harmonic from Eq. (12) would also have entered into the averaged result. In the experiment, however, the second harmonic is absent to a rather good accuracy (see Figs.4(*a*) and 4(*b*)), showing that the choice Eq.(13) is fortunate. One should note, however, that – although in numerous recent polarization-resolved studies of single QDs the long axes of the QDs do tend to align along the {110}-type directions – examples of clearly different orientations of the QDs are also routinely observed.

In closing, as the angular distribution of the QDs is now reconstructed, we shall also briefly discuss the value of the built-in polarization. The measured amplitude of this polarization amounts to $\rho_0 \sim 0.6\%$ (Fig.4(*d*)). Because of the angular scatter of the axes of QD elongation, this value must be much smaller than the typical value of the built-in polarization of a single QD, $\rho_0^{SQD}$. It is easy to show that for the angular distribution given by Eq. (13), $\rho_0 = \rho_0^{SQD} \cdot a/(2 + 2a + 2b)$, which yields $\rho_0^{SQD} \sim 3.5\%$ for the deduced values *a*, *b* and $\rho_0$. This is notably smaller than the typical values $\sim 30\%$ reported for trion states in singly charged QDs.[16,19]

**Conclusions**

In the present paper the magneto-optical inter-polarization conversion by an ensemble of CdSe/ZnSe quantum dots was investigated both experimentally and theoretically. We studied various types of polarization responses of the QDs as a function of applied magnetic field and of the orientation of the sample. Apart from the already known patterns of polarization behavior we report the first observation of a 90-degree anisotropy of the optical alignment and of the $LL'$ conversion at zero magnetic field. At the same time, to the best of our experimental accuracy, no $LC$ or $CL$ conversions were found at zero field.

The set of experimental data was analyzed using two scenarios of spin evolution of the exciton: the two-step model and the rudimentary one-step model. It was found that, although the one-step model can formally reproduce many of the experimental observations, it leads to an unclear value of the characteristic lifetime $\tau$, and it fails to account for the absence of the zero-field $LC$ and $CL$ conversions. In using the two-step model with a long-lived ground exciton state, one does not encounter these difficulties. Although we were unable to make a precise determination of the value of the anisotropic exchange splitting of the excited exciton state by using the two-step analysis (the results of simulation are rather tolerant to the choice of this parameter), this value is apparently several times larger than that of the ground state, and the lifetime of the excited state is in the sub-picosecond range.

To explain the experimental data with either of the models, one has to introduce an in-plane angular scatter of the elongation axes of the QDs. We found that a description of the angular scatter by the distribution function given by Eq.(13) provides very satisfactory results, and the parameters $a$ and $b$ determining this distribution could thus be reliably evaluated from the comparison of experimental and simulated polarization conversion results. No significant difference was found in the $a$ and $b$ values obtained via the one-step and two-step models. Thus angular distribution reconstructed by both models (see Fig. 5) shows a conspicuous tendency for the QDs to align with the $\{110\}$-type directions, with a further preference for the $[110]$ direction as compared to $[1\bar{1}0]$.

In the QD sample studied here, the angular scatter of the QDs turned out to be quite strong, thus obscuring the interesting consequences of the two-step scheme of the conversions. For example, because of experimental limitations we were not able to reliably measure an even (in the magnetic field) W-shaped linear-to-circular polarization conversion at $\varphi = \pi/4$ (see Eq. (15)), a pattern which is predicted but has never been reported. However, the agreement of our present data with the details of the two-step model holds out the promise that, with further refinement of experimental technique, semiconductor nanoheterostructures with more regular in-plane anisotropy will provide an opportunity for observing this effect.

**Acknowledgments**

The support of the Russian Foundation for Basic Research (Grants 06-02-16676 and 06-02-16603); National Science Foundation Grant DMR06-03762; and the Korea Research Foundation Grant KRF-2004-005-C00068 is gratefully acknowledged.

**APPENDIX**

Here we present the results of calculations of the polarization responses within the one-step model of polarization conversion. Instead of Eqs. (11) and (12), this model gives

$$\begin{pmatrix} CL \\ CL' \\ CC \end{pmatrix} = \begin{pmatrix} \Omega_{ex1}\Omega_B \tau^2 \cos 2\phi - \Omega_{ex1}\tau \sin 2\phi \\ \Omega_{ex1}\Omega_B \tau^2 \sin 2\phi + \Omega_{ex1}\tau \cos 2\phi \\ 1 + \Omega_B^2 \tau^2 \end{pmatrix} \frac{1}{1+(\Omega_{ex1}^2+\Omega_B^2)\tau^2}, \quad (A1)$$

$$\begin{pmatrix} LL \\ LL' \\ LC \end{pmatrix} = \begin{pmatrix} 1 + \Omega_{ex1}^2 \tau^2 \cos^2 2\phi \\ \Omega_{ex1}^2 \tau^2 \sin 2\phi \cos 2\phi - \Omega_B \tau \\ \Omega_{ex1}\Omega_B \tau^2 \cos 2\phi + \Omega_{ex1}\tau \sin 2\phi \end{pmatrix} \frac{1}{1+(\Omega_{ex1}^2+\Omega_B^2)\tau^2}, \quad (A2)$$

where $\Omega_{ex1}$ denotes the value of the exchange field in the only state of the one-step model. The above equations contain some qualitative dissimilarities compared to the two-step model presented in the main text of this paper. Equations (A1) and (A2) indicate that optical orientation $CC$ and conversions $LC$ and $CL(CL')$ are possible at zero magnetic field (although they will be small for $\Omega_{ex1}\tau \gg 1$). According to Eq. (A2) the optical alignment $LL$ is a strictly an even function of the magnetic field, while Eq. (12) of the two-step approach allows a component of the effect that is an odd function of the field. Furthermore, Eq. (11) of the two-step model indicates that for $C$ excitation the linearly polarized component of the response will have the same orientation at any value of the magnetic field, while Eq. (A1) predicts a rotation of the plane of polarization as the field is varied. Finally, the pattern of the field dependence of the $LC$ conversion at $\phi = \pi/4$ is distinctly different in the two models.

Averaging of Eqs. (A1), (A2) with the angular distribution given by Eq. (13) leads to the following results (cf. with Eqs. (14), (15)):

$$\begin{pmatrix} CL \\ CL' \\ CC \end{pmatrix} = \begin{pmatrix} \Omega_{ex1}\Omega_B \tau^2 \frac{a}{1+a+b}\frac{\cos 2\varphi}{2} - \Omega_{ex1}\tau \frac{a}{1+a+b}\frac{\sin 2\varphi}{2} \\ \Omega_{ex1}\Omega_B \tau^2 \frac{a}{1+a+b}\frac{\sin 2\varphi}{2} + \Omega_{ex1}\tau \frac{a}{1+a+b}\frac{\cos 2\varphi}{2} \\ 1 + \Omega_B^2 \tau^2 \end{pmatrix} \frac{1}{1+(\Omega_{ex1}^2+\Omega_B^2)\tau^2}, \quad (A3)$$

$$\begin{pmatrix} LL \\ LL' \\ LC \end{pmatrix} = \begin{pmatrix} 1 + \Omega_{ex1}^2 \tau^2 \left( \dfrac{1}{2} + \dfrac{b}{1+a+b} \dfrac{\cos 4\varphi}{4} \right) \\ \Omega_{ex1}^2 \tau^2 \dfrac{b}{1+a+b} \dfrac{\sin 4\varphi}{4} - \Omega_B \tau \\ \Omega_{ex1} \Omega_B \tau^2 \dfrac{a}{1+a+b} \dfrac{\cos 2\varphi}{2} + \Omega_{ex1} \tau \dfrac{a}{1+a+b} \dfrac{\sin 2\varphi}{2} \end{pmatrix} \dfrac{1}{1 + (\Omega_{ex1}^2 + \Omega_B^2)\tau^2}. \quad (A4)$$

The effect of relaxation of the pseudospin would have resulted, as in the formulas of the main text, in the form of the common prefactor $T_s / \tau_{rad}$ on the right-hand side of Eqs. (A1) – (A4). In addition, when relaxation is present, $\tau$ in Eqs. (A1) – (A4) should be understood as representing the pseudospin relaxation time ($T_s$), rather than the exciton lifetime ($\tau_{rad}$).

As a final note, we recall that not every dissimilarity of Eqs. (A1) – (A4) compared to the corresponding formulas in the main text should be ascribed just to the number of steps (one or two) in the scheme of the pseudospin evolution. In the main text, in order to keep the results more compact (and in view of the experimental conditions used in this paper), in our two-step model we assumed a long precession of the pseudospin in the lower level ($\omega_{ex} \tau_{rad} \gg 1$, see Eq. (10)). In contrast, Eqs. (A1) – (A4) were obtained with no limitations imposed on the of value $\Omega_{ex1} \tau$.

**FIGURE CAPTIONS**

Fig. 1. Schematic depiction of two models of spin evolution. (a) The two-step model. Here the precession of the pseudospin in the excited state of the exciton is interrupted by the escape to the ground state of the exciton (with a characteristic lifetime $\tau$). The long-lasting precession in the ground state is followed by the recombination of the exciton (lifetime $\tau_{rad}$). (b) The one-step model. Here the pseudospin precesses in the only state of the model, but no limitations are imposed on the duration of the precession.

Fig. 2. PL spectrum of the CdSe/ZnSe QD sample, and a corresponding spectrum of optical alignment in the $LL$ configuration.

Fig. 3. Various types of the polarization response as a function of applied magnetic field. Points – experimental data; solid curves – calculations using the two-step model (Eqs. (14) and (15)); dashed curves (sometimes merging with solid lines) – calculations using the one-step model (Eqs. (A3) and (A4)). For values of parameters used in the calculations see text.

Fig. 4. Zero-field angular dependences of (a) optical alignment, (b) linear-to-linear conversion, and (c) linear-to-circular conversion. Types of curves as in Fig.3. Panel (d) shows angular dependence of the built-in linear polarization, which is not a polarization response, and was observed with non-polarized excitation. Curves in part (d) show a simple second-harmonic ($\cos 2\varphi$) fit.

Fig. 5. Reconstructed distribution of QD elongations over the direction $\psi$ (where $\psi = 0$ indicates that the long axis of a QD is parallel to the [110] direction). The distribution is calculated using Eq.(13) with values $a = 0.8$, $b = 0.5$, as deduced from experiment.

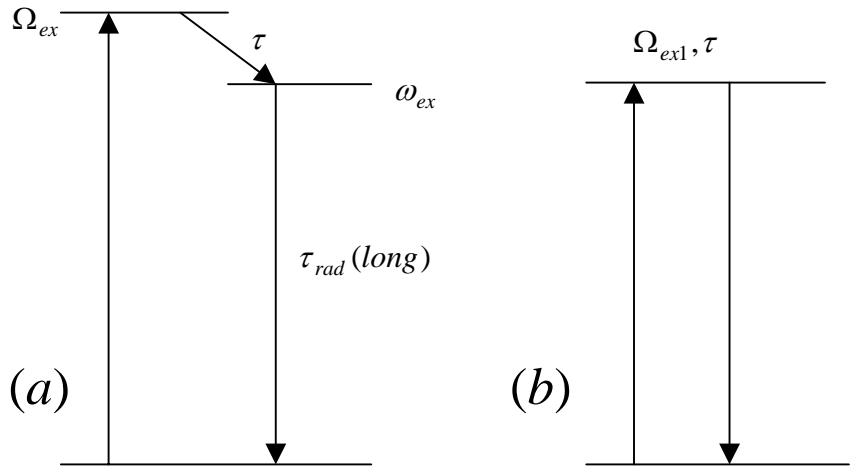

Fig. 1. Schematic depiction of two models of spin evolution. (a) The two-step model. Here the precession of the pseudospin in the excited state of the exciton is interrupted by the escape to the ground state of the exciton (with a characteristic lifetime $\tau$). The long-lasting precession in the ground state is followed by the recombination of the exciton (lifetime $\tau_{rad}$). (b) The one-step model. Here the pseudospin precesses in the only state of the model, but no limitations are imposed on the duration of the precession.

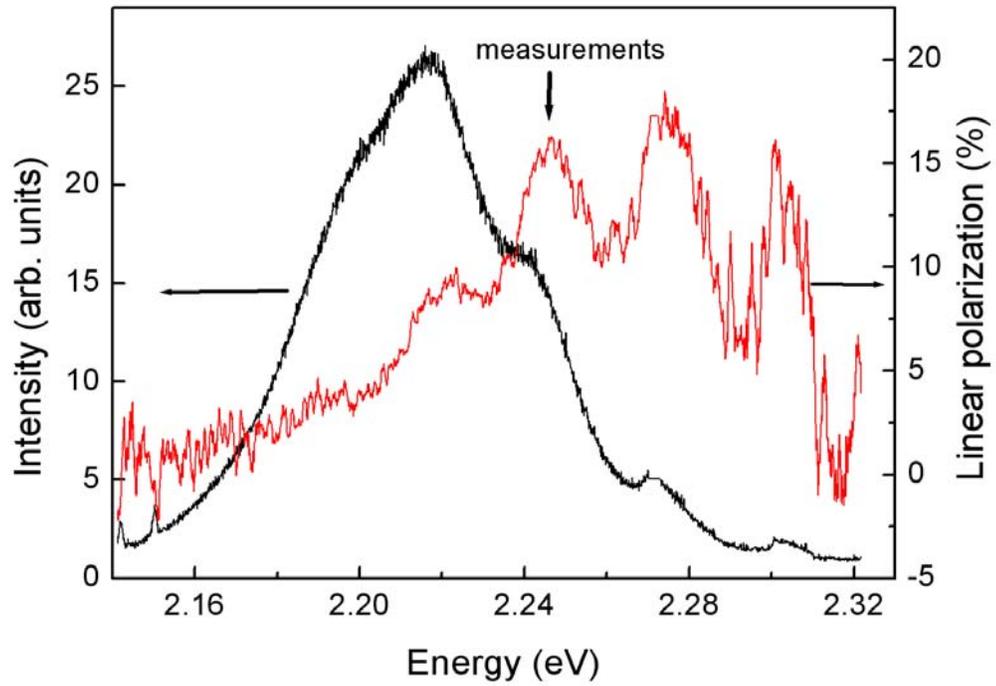

Fig. 2. PL spectrum of the CdSe/ZnSe QD sample, and a corresponding spectrum of optical alignment in the *LL* configuration.

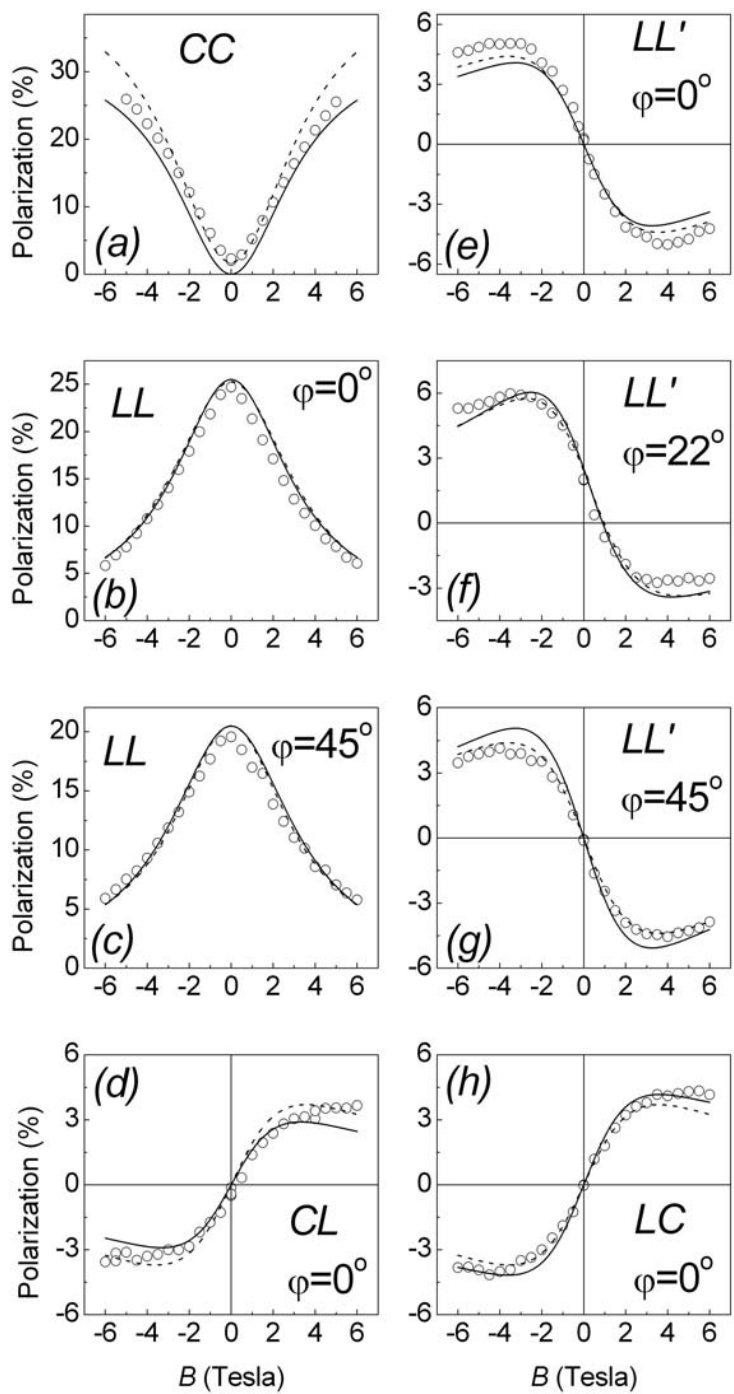

Fig. 3. Various types of the polarization response as a function of applied magnetic field. Points – experimental data; solid curves – calculations using the two-step model (Eqs. (14) and (15)); dashed curves (sometimes merging with solid lines) – calculations using the one-step model (Eqs. (A3) and (A4)). For values of parameters used in the calculations see text.

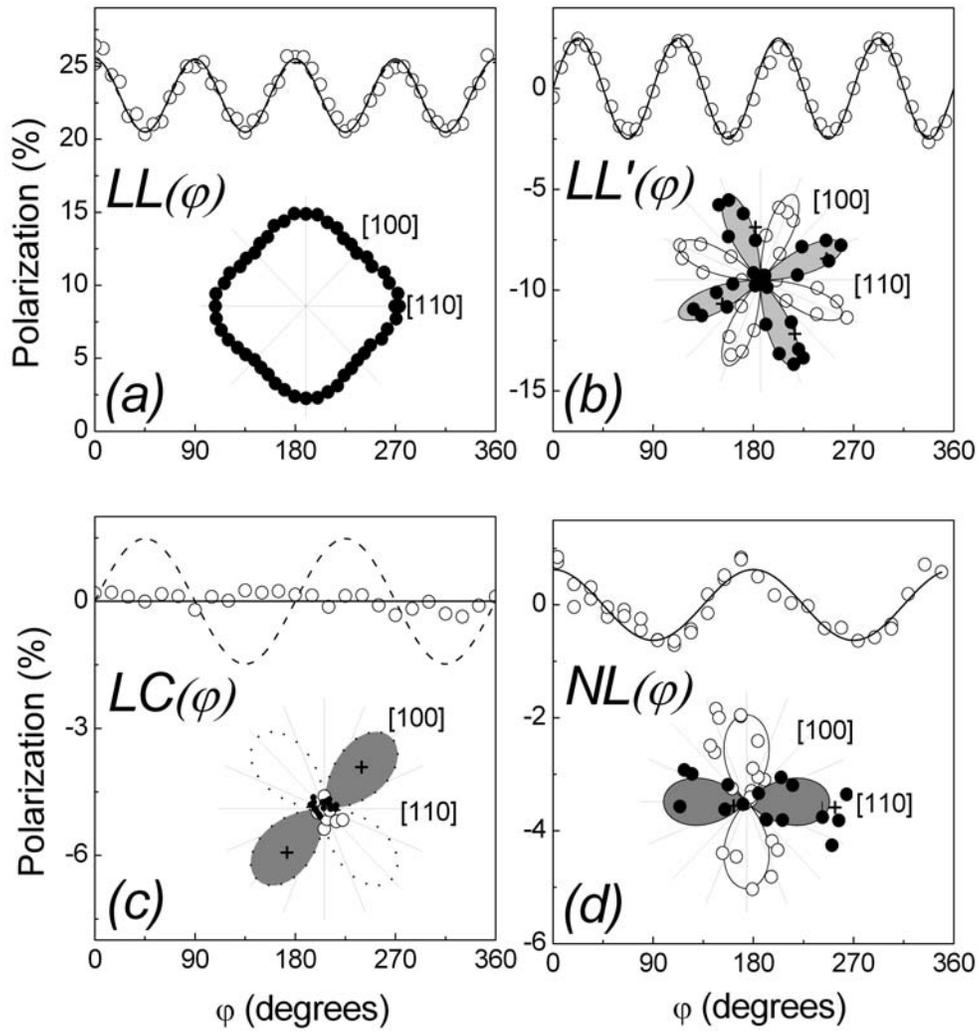

Fig. 4. Zero-field angular dependences of (a) optical alignment, (b) linear-to-linear conversion, and (c) linear-to-circular conversion. Types of curves as in Fig.3. Panel (d) shows angular dependence of the built-in linear polarization, which is not a polarization response, and was observed with non-polarized excitation. Curves in part (d) show a simple second-harmonic ($\cos 2\varphi$) fit.

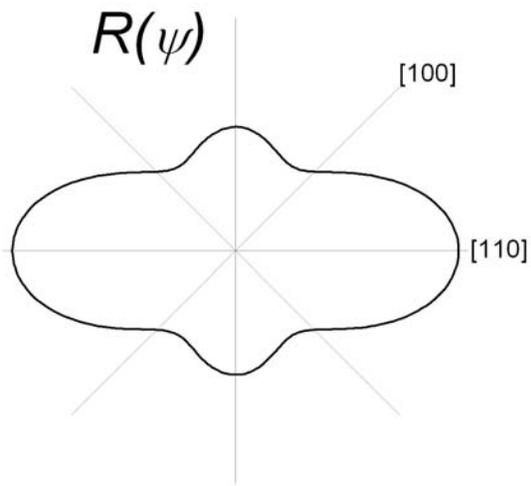

Fig. 5. Reconstructed distribution of QD elongations over the direction $\psi$ (where $\psi = 0$ indicates that the long axis of a QD is parallel to the [110] direction). The distribution is calculated using Eq.(13) with values $a = 0.8$, $b = 0.5$, as deduced from experiment.